# MindSculpt: Using a Brain–Computer Interface to Enable Designers to Create Diverse Geometries by Thinking

Qi Yang[1], Jesus G. Cruz-Garza[2], Saleh Kalantari[*]

**Abstract**

MindSculpt enables users to generate a wide range of hybrid geometries in Grasshopper in real-time simply by thinking about those geometries. This design tool combines a brain–computer interface (BCI) with the parametric design platform Grasshopper, creating an intuitive design workflow that shortens the latency between ideation and implementation compared to traditional computer-aided design tools based on mouse-and-keyboard paradigms. The project arises from transdisciplinary research between neuroscience and architecture, with the goal of building a cyber-human collaborative tool that is capable of leveraging the complex and fluid nature of thinking in the design process. MindSculpt applies a supervised machine-learning approach, based on the support vector machine model (SVM), to identify patterns of brain-waves that occur in EEG data when participants mentally rotate four different solid geometries. The researchers tested MindSculpt with participants who had no prior experience in design, and found that the tool was enjoyable to use and could contribute to design ideation and artistic endeavors.

*Keywords:* *Brain Computer Interface, Design Tool, Intuitive Interaction, Design Cognition, Artificial Intelligence, Machine Learning*

## 1.     Introduction

Abraham Maslow is credited with formulating the "law of the instrument," which is commonly summed up with the quotation: "If all you have is a hammer, everything looks like a nail" (Maslow 1966, p.15). This phrase reminds us that the tools we use in design will inevitably affect our perception and workflow, and thus contribute to shaping the resulting design products. In the age before powerful computers, designers such as Antoni Gaudí and Frei Otto developed physical prototyping tools to help them better visualize the complex forms taking shape in their minds. The primary purpose of these tools was to provide timely feedback so that the designers could evaluate their visions and make improvements. As the computer era advanced, design workflows based on the WIMP paradigm (Widows, Icons, Menus, and Pointers) became the new standard (Caetano et al. 2020; Shanker et al. 2014). These tools were meant to enable a more intuitive technological interface for people without programming backgrounds (Sutherland 1964). However, multiple empirical studies have indicated that using WIMP-based computer-aided design tools can impede design ideation and the conceptual design process. Compared to old-fashioned sketching, WIMP-based design has been shown to discourage the generation of diverse ideas (Stones et al. 2007; Alcaide-Marzal et al. 2013), to mislead designers to focus excessively on details and neglect the big picture. (Ibrahim et al. 2010; Charlesworth and Chris 2007), and promote premature design fixation and constrain designers' capabilities due to the unintuitive interface (Robertson and Radcliffe 2009).

Some researchers have put forth plausible theories and principles for tools that can better support design creativity (Hutchins et al. 1985; Shneiderman et al. 2006). For example, according to Hutchins's direct manipulation theory design tools should seek to minimize the "gulf of execution," which is the mental gap or delay that occurs when computer users have to translate their intentions into linear and sequential machine-understandable commands. A significant gulf of execution will lead to high latency between idea-generation and realization, interrupting the intuitive process of visualization and design feedback. One of the main problems with WIMP-based design tools is that they have relatively significant gulfs of execution, particularly for users who are not yet fully fluent in the relevant software and input technologies.

To address the deficiencies of WIMP-based computer interfaces during the design ideation phase, we developed MindSculpt. This new design tool directly translates the user's imagination of different forms into high-fidelity, real-time digital prototypes in common design software platforms, without the need for pointing-and-clicking (Figure 1). MindSculpt is based on a machine learning algorithm that has been trained to identify brain signals associated with the user's visualization of different solid geometries. This technology is a work in progress, and the output is currently quite fluid and, in some cases, difficult to control, mirroring the complex nature of thinking itself. Overall, however the feedback from our test participants has been quite positive, indicating that MindSculpt has significant potential for application in design ideation.

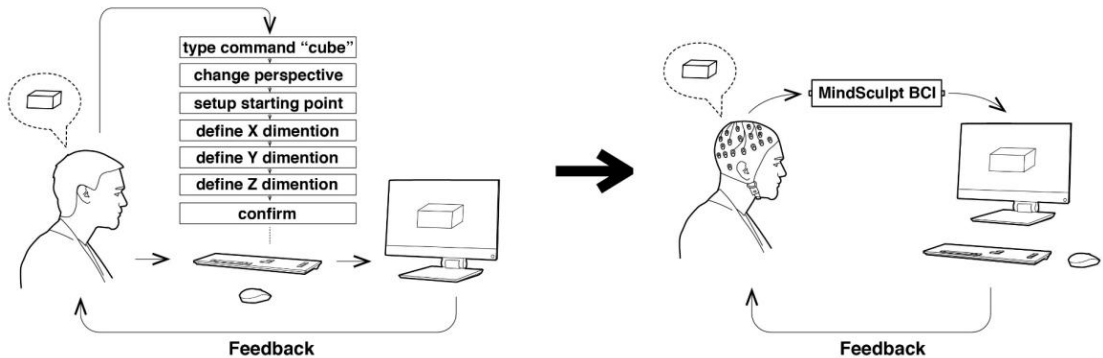

**Figure 1. Traditional CAD Tool Feedback Loop v.s. the MindSculpt Feedback Loop.**

## 2.　　Background

Brain–computer interfaces (BCI) are capable of translating a users' brain activities into messages or commands (Lotte et al. 2018). Previous studies have demonstrated the possibility of classifying EEG signals associated with the mental visualization of complex geometries and shapes (Esfahani et al. 2012; Rai and Akshay 2016). Furthermore, the mental rotation of objects has become a common task used in the BCI field to demonstrate the technology's potential for good binary classification scenarios (Friedrich et al. 2013; Jeunet et al. 2015). Using mental rotation may accommodate a wider range of task difficulties and offer users more flexibility when using BCI (Gardony et al. 2017). Although the majority of BCI applications are oriented toward clinical scenarios, especially for physically disabled people (Abiri et al. 2019), the use of BCI technology in art and design has also been explored (Nijholt 2019, p.104-111). For example, in an early study researchers used brain activity to control a group of robots in an art installation (Ulrike 1991). Other studies enabled users to create digital paintings, control panel movement, or build digital models using brain signals (Barsan-Pipu 2019; Kübler et al. 2019; Shanker et al. 2014; Todd et al. 2019; Wulff-Abramsson et al. 2019; Herrera-Arcos et al. 2019). Kovacevic and colleagues (2015) acquired brain signals from 20 participants simultaneously and used those data to manipulate an ambient room interior as a form of immersive art.

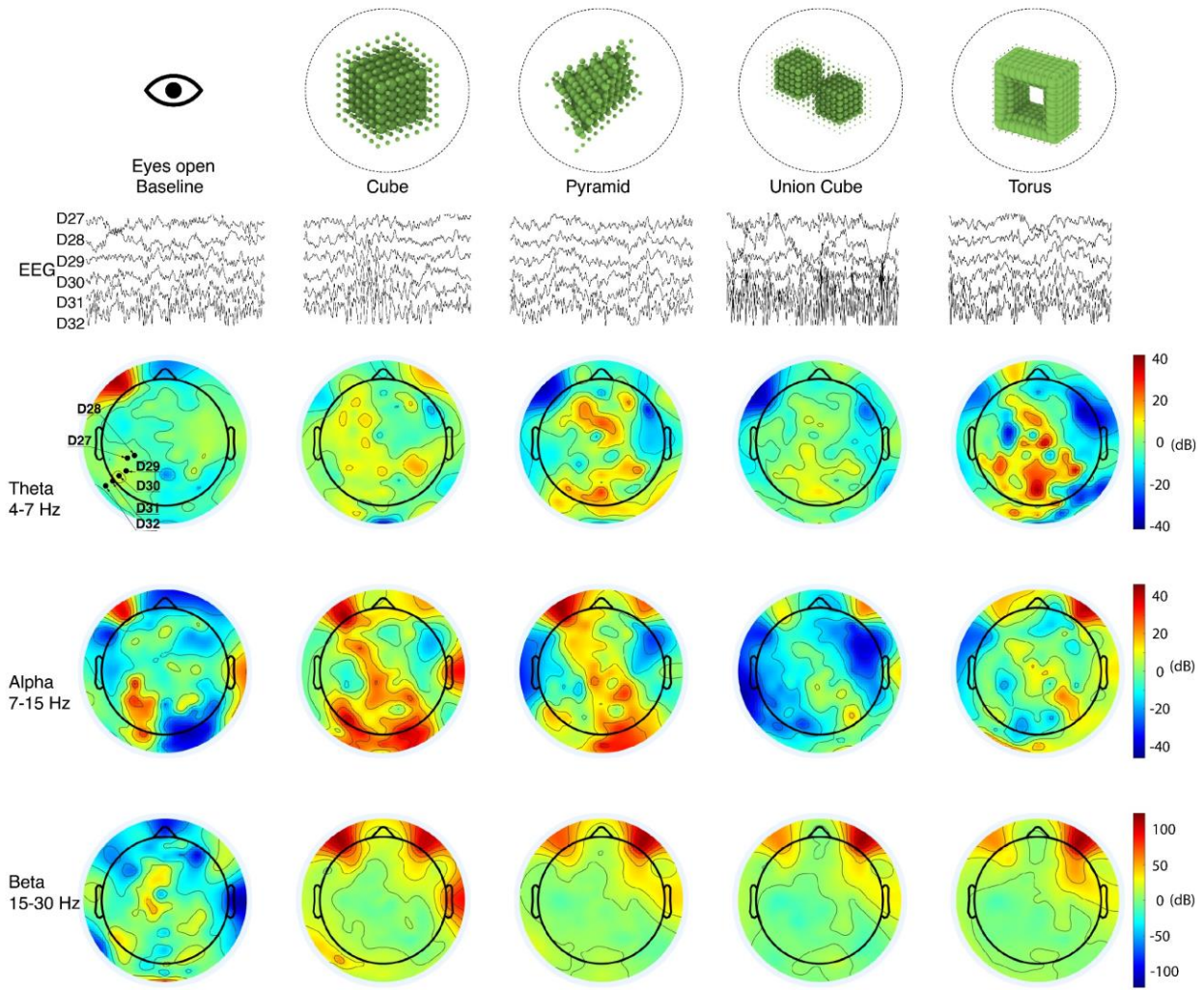

**Figure 2. Different Brain Patterns Associated with the Mental Rotation of Each Geometry for Participant 1.**

In many of those studies, the BCI applications created intriguing external visualizations of brain signals and have the capability to customize the relationship between the user's mental imagery and the output. However, the type of brain activity linked to various BCI outputs was often tangential or even counterintuitive. One system has mapped motor activities such as eyeblinks to commands such as "draw a circle" (Shanker et al. 2014). One system made icons flicker to decode where the user attended to (Kübler et al. 2019). Another exploration used brain activity to evaluate and influence an autonomously developing design rather than to exert active, dynamic control (Barsan-Pipu 2019). Here, we provide an experience to the user where a BCI is used to create an intuitive and instant feedback loop between the ideation of geometric forms and their external representation.

The MindSculpt project used a machine learning approach to identify neural signals associated to the mental visualization of different solid geometries, and then represented those same geometries on the screen via common design software programs (Figure 2). Discrete classification paradigms seemed insufficient to satisfy the need for design exploration during the ideation and conceptual design phases. In Mindsculpt, the dynamic shapes generated by the users were created as a combination of the component solid shapes that the BCI was trained for; harnessing the capabilities of discrete classification paradigms (i.e. training the BCI on four shapes), while providing an evolving output (Figure 3).

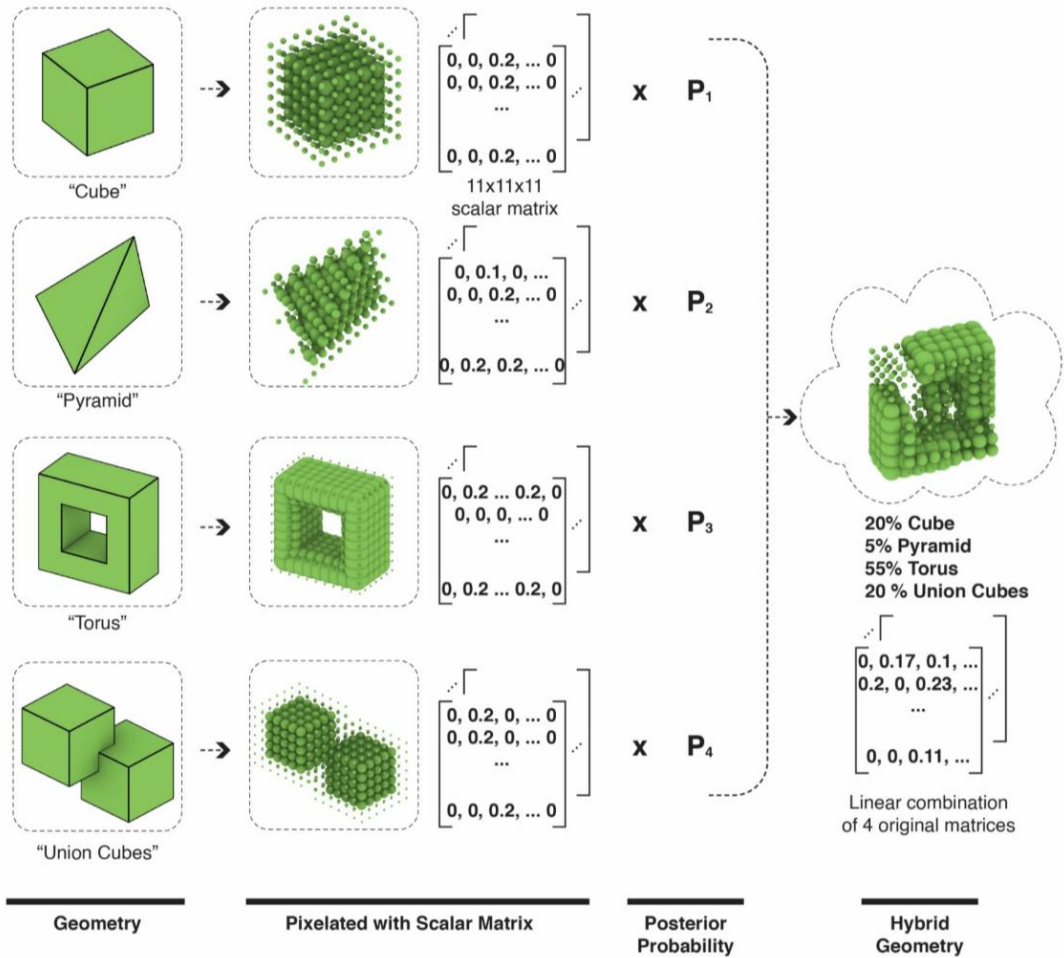

**Figure 3. Geometry Generation Using Posterior Probability Allows for the Creation of Hybrid Figures.**

The approach that we adopted for MindSculpt used the posterior probability score as the output of the machine learning model, and pixelated the 3D geometries based on the probability score. The posterior probability was the probability assigned to a new data sample to belong to each of the class labels that the model was trained for. As a result, MindSculpt was able to provide a wide spectrum of complex geometries beyond the four basic shapes that were used to train the machine-learning model. Thus, for example, when the user was hesitating between a cube and a pyramid during the ideation process, MindSculpt would fluidly generate a novel, hybrid geometry that was in-between a cube and a pyramid (Figure 3).

## 3.    Method

During the MindSculpt project the researchers first determined the best approach to creating the BCI, and then conducted a pilot study with non-designer participants to test the system.

### 3.1.    Hardware and Software Setup

The current set-up of MindSculpt included an EEG headset, Openvibe (Renard et al. 2010), Matlab and Grasshopper (Figure 4). EEG data were collected using a non-invasive, 128-channel, gel-based Actiview System (BioSemi Inc., Amsterdam, Netherlands) with Ag/AgCl active electrodes. Openvibe was an open-source BCI platform that could acquire real-time EEG data and transfer the results to Matlab through the Lab Streaming Layer (Kothe et al. 2014). Matlab was used to pre-process the EEG data, train the participant-specific machine-learning model, predict the subsequent geometries that users visualized, and send the target class posterior probability scores to Grasshopper in the User Datagram Protocol (UDP) format. The capabilities of Grasshopper allowed us to script different geometry generation algorithms and visualize the outcomes to the users in real-time.

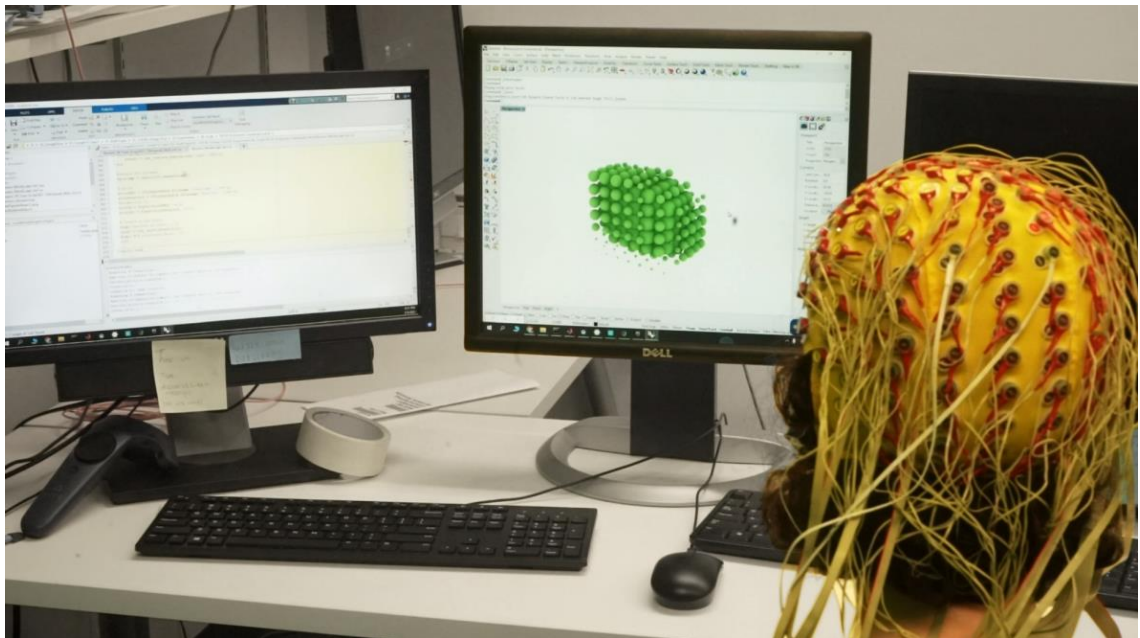

**Figure 4. Experiment Scenario of One Participant Using MindSculpt to Ideate.**

3.2.    MindSculpt Workflow

Using MindSculpt was a three-stage process (Figure 5): first, participants were fitted with the EEG cap and received instructions about mental rotation; then they underwent a 5-minute training session and completed an additional 5-minute machine-learning session so that the BCI could evaluate their personal neurological responses; and then finally they could freely use the tool in real-time. Previous studies had shown that BCI self-regulation was learning psychological factors, including cognitive states experienced by the users that session e.g. motivation, concentration, flow; as well as external factors as the time of measurement (Roc et al. 2020). These factors could affect the EEG-data mental imagery representation of the same abstract object at different sessions, for distinct participants. Therefore, we built a personalized ML model for each participant (Shah et al. 2021; Wang et al. 2021). The training session was meant to calibrate the BCI for different users and optimize the performance.

During this session, the participant viewed a video in which each of the four pre-selected geometries appeared on the computer screen for 10 seconds in a randomized sequence. Those geometries were a cube, a pyramid, a square torus, and two cubes combined diagonally ("union cubes"). When any geometry appeared, users were asked to perform a rotation of the image in their minds, which helped to create a strong mental visualization of the figure (Friedrich et al. 2013). This process continued until each figure had been shown five times.

As the participants viewed and visualized these solid geometries, EEG data were recorded at 256 Hz. At the end of the training session these data were processed and put through the machine-learning pipeline in Matlab (Figure 5). Raw EEG data were preprocessed in order to bandpass the preferred frequency band (1~40 Hz), and remove bad channels with poor contact with the scalp. We applied artifact subspace reconstruction (Mullen et al. 2013) to remove noisy channels and artifactual power bursts (sd threshold = 15). Then, we visually inspected and removed EEG spikes caused by motion artifacts such as blinking, clenching, and body movements because they would cause confusion for the ML model to classify mental rotation of different geometries.

For each of the 128 EEG channels, we extracted the power in three frequency bands, in windows of 2s of data in steps of 0.5s. The frequency bands selected were the theta (4-7 Hz), alpha (7-15 Hz), and beta (15-30 Hz).

We used Minimum Redundancy Maximum Relevancy (mRMR) (Peng et al. 2005) algorithm to select the most salient features out of 384 features (Figure 7). We defined the minimum number of selected features as 16 to increase the robustness of the subsequent real-time performance because the data of any channel could be affected by unpredicted events such as fierce movements from the user in real time. MindSculpt then used the data to train support vector machine models (SVM), creating a unique participant-specific model based on the brain activities associated with each of the four geometries.

At this point, the participant could begin using MindSculpt to visualize hybrid shapes in real-time. During our pilot study, participants were asked to visualize two-class combinations of the sample geometries. These combinations included [cube + pyramid], [union cube + torus], and [pyramid + union cube]. The participants were also asked to visualize one four-class combination, in which all of the sample geometries were combined to create a final shape. They were told to create these hybrid geometries however they wished, but encouraged to consider mental rotations and mergers of the sample shapes, and to work on each hybrid geometry for at least 3 minutes. The trained machine-learning model, in turn, would assign weights to those four pre-defined geometries that were directly proportional to the posterior probability that a data sample belonged to each class. The hybrid geometry derived from the EEG data was represented on a computer screen by Grasshopper in real-time (Figure 5). Users were able to press a button to pause the reception of brain data and save the displayed geometry if they liked the design. They were then able to use the same button to resume the data reception and real-time visualization to continue exploring their mental control of the system. After the experiment, the researchers conducted a brief, semi-structured debriefing interview to obtain each participants' feedback about the MindSculpt system performance.

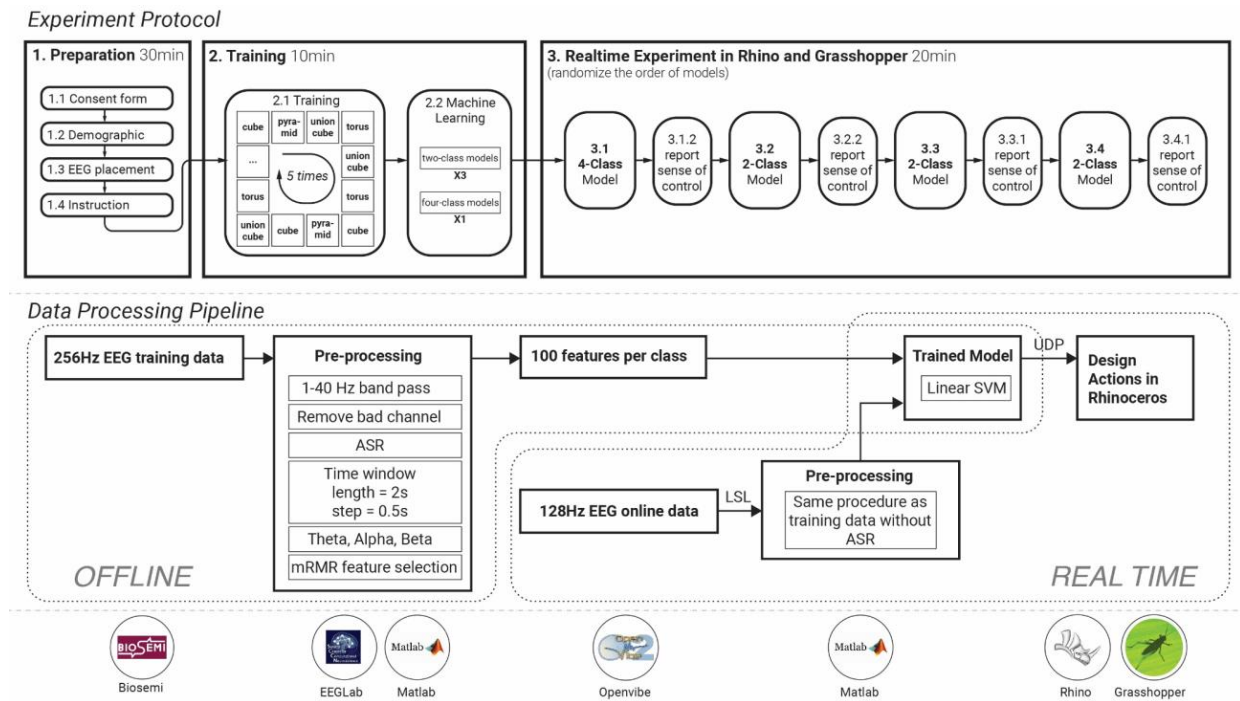

**Figure 5.  Experiment Protocol and Data Processing Pipeline of MindSculpt.**

### 3.3.    Participants

For the feasibility pilot study, we recruited seven participants using a convenience sampling method (word-of-mouth and announcements on departmental e-mail lists). None of the participants had prior experience working or studying in design fields, nor did they have prior BCI experiences. All of them were undergraduate university students. Each participant gave informed written consent before participating in the experiment, and the overall study protocol was approved by the Institutional Review Board prior to the start of research activities. All of the experiment sessions took place at the same physical location.

## 4.    Results

For the seven participants in the pilot study, we reached 82% mean classification accuracy for the training set 2-class models (SD = 4%, n = 7), and 78% mean off-line validation accuracy (SD = 3%, n = 7) by iteratively using 4 trials to train a Linear SVM classifier and 1 trial as the validation set (Figure 6).

We also asked participants to self-report their sense of control over the represented hybrid geometry on a scale from 1 (low) to 10 (high). 5 out of 7 participants' data were collected. Although some participants reported a high sense of control when using MindSculpt, we observed a wide range of perceived performance for the 2-class models (min = 1.67, max = 7.3, M = 4.93, SD = 2.29, n = 5). The high variance of self-reported performance in this study matched the observations from previous BCI research (Jeunet et al. 2015).

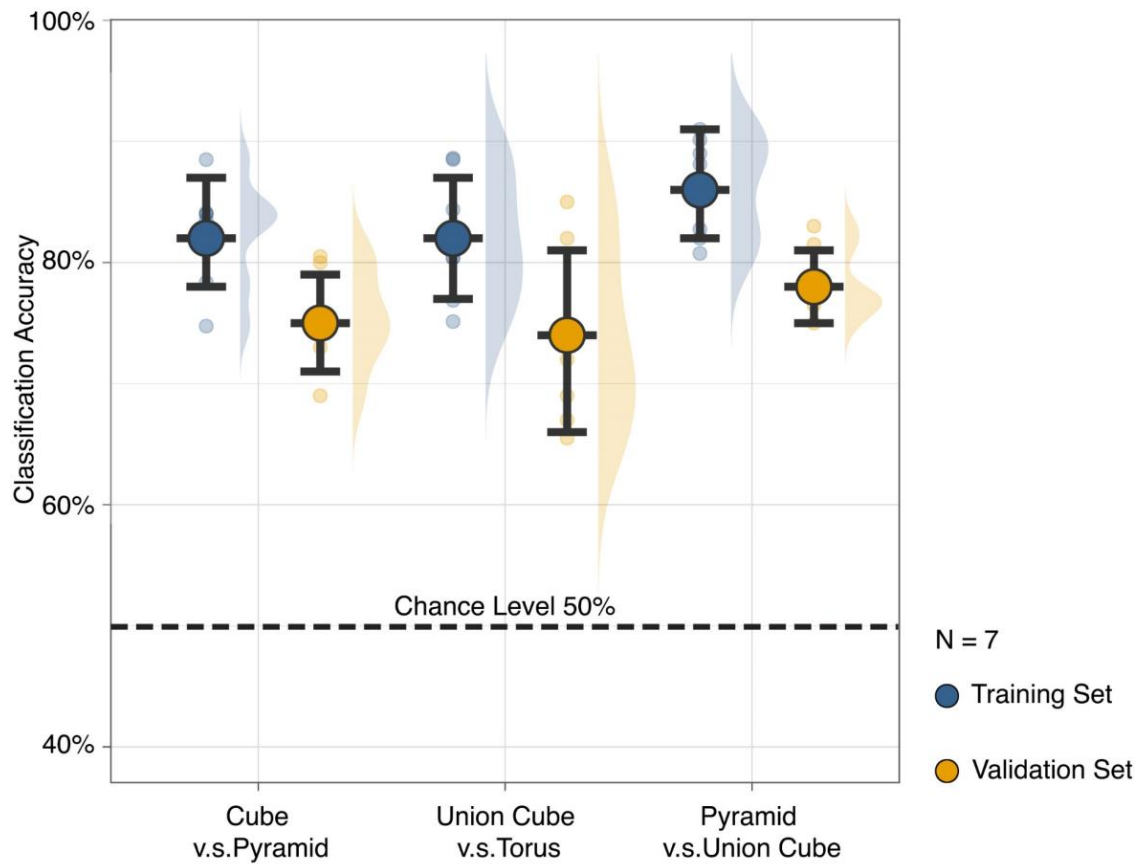

**Figure 6.  Classification and Validation Accuracy for Three 2-class Models.**

More alpha power was observed in frontal and posterior regions of the scalp for the simple geometries (cube, pyramid) and suppression for more complex geometries (union, torus) (Figure 2). The observations resonated with previous findings (Michel et al. 1994; Riečanský et al. 2010; Gardony et al. 2017). In terms of selected EEG features, 23 features were selected on average for all 2-class models (SD = 6.83, n = 21). The feature selected showed that the validation accuracy reached a plateau (or decreased) after 17 to 23 features selected. More features would potentially impair real time performance. Selected features for each model varied among different participants (Figure 7), but comprised a network of frontal, central, and parietal regions.

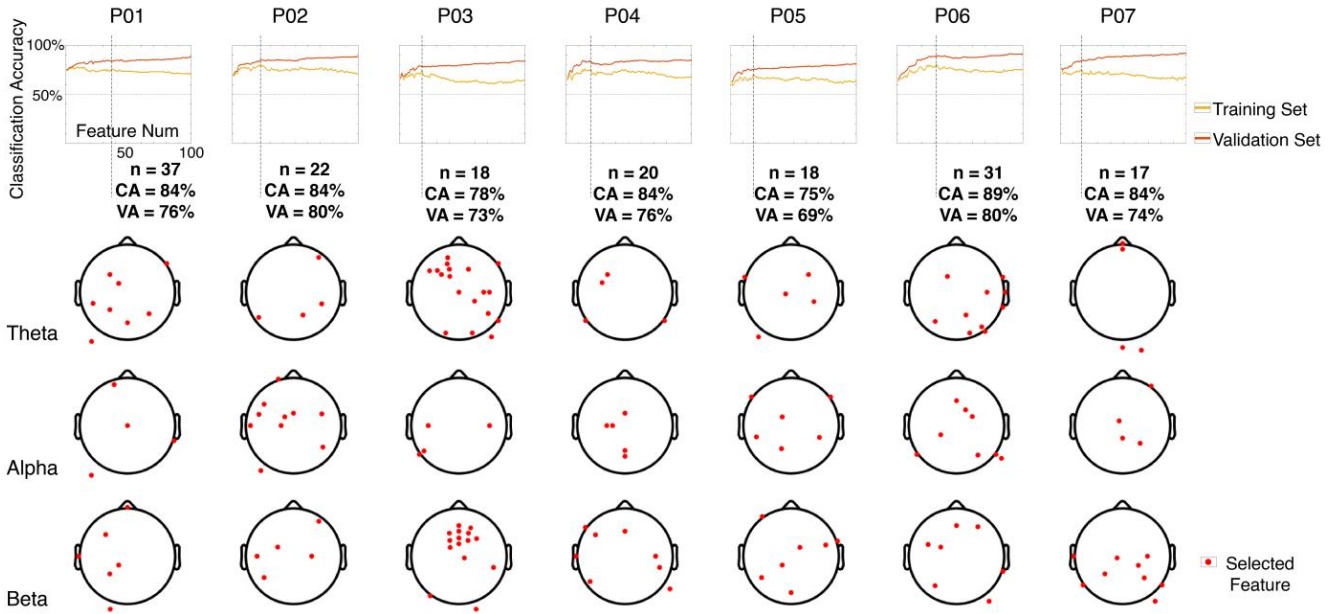

**Figure 7. Feature Selection for All Participants Based on mRMR of Cube v.s. Pyramid Model**.

The most relevant features for classification, as obtained with mRMR, showed distinct patterns of channel relevance when participants were mental rotating geometries with different complexity (Figure 7). In the alpha band (8-12 Hz), central regions (close to motor regions) showed consistently most relevance and least redundancy to predict the mental imagery; although the spatial location of the electrodes was broad. The theta (4-7 Hz) and Beta (15-30 Hz) frequency bands also provided relevant information for classification performance, including relevant channels in midline frontal and parietal areas (Gardony et al. 2017).

The user testing with these seven participants produced inspiring feedback. During the experiment, we noticed that almost all of the participants were willing to spend much more time than the required 3 minutes to explore each hybrid shape. Three participants reported that they were motivated to try different thoughts other than mere mental rotation and combination to generate geometries. For example, one participant explored thinking about rectangular objects and tested if a cube could be stretched accordingly.

In the interviews, all participants described the tool as "inventive" and "interesting." They said that they felt motivated to explore different shapes, and reported that the instant real-time feedback was helpful for their thinking process. However, one participant mentioned that they had difficulty concentrating on their internal visualization while also watching the changing real-time feedback on the screen. One participant indicated that he preferred to close his eyes and think, then open them briefly to check the outcomes. Five participants reported difficulty and high mental workload in continuously maintaining focus on the geometric shapes. When asked explicitly about the relationship between the MindSculpt BCI and conventional WIMP-based computer-aided design, three participants indicated that we should use both techniques in the future. When asked about the limitations of MindSculpt, the most common concern was the BCI's lack of accuracy in clearly capturing the user's mental model on the screen. When the EEG-based representation failed to match participants' intentions for what seemed like a long period of time, three participants experienced feelings of confusion, frustration, or annoyance. We even observed one participant self-blamed themselves for not thinking hard enough though it was not the participant's fault.

## 5.    Discussion and Conclusion

In this paper, we introduce a new modality of interaction, brain-computer interface to the architectural design field and explores its potentials in design ideation. Distinct from previous BCI applications (Shanker et al. 2014; Kübler et al. 2019), MindSculpt is the first BCI application for designers in the architectural form-making ideation phase that directly matches users' intentions with the corresponding visual feedback, with aims to minimize the gulf of execution (Hutchins et al. 1985). The study involves recording participants' brain activity using a 128-channel EEG headset and developing a machine-language classifier to reflect the users' mental rotation of solid geometries. Adding to the previous study that uses mental imagery tasks to classify different geometries (Esfahani et al. 2012), MindSculpt demonstrates a novel alternative to classify geometries with different complexity by asking user to conduct mental rotation tasks.

To test the performance of the interface, a feasibility study with 7 participants is conducted. The results indicates that this approach to human–computer interaction has significant potential, though the reliability of the system degenerates when users are confused by the instruction and not sure what to imagine, or distracted during the training session.

The advantage of using a BCI for designers is that it can create a more intuitive, real-time mechanism of computer-aided design to facilitate cognitive processing, particularly during the design ideation stages of product development. Since essentially all other current approaches to human–computer interaction is restricted to the WIMP model (Kerous, Skola, & Liarokapis, 2018), a BCI-based approach such as MindSculpt has the potential to revolutionize mainstream design practice. MindSculpt aims to shorten the feedback loop from design ideation to design implementation, so that practitioners can get instant, high-fidelity feedback and external visualization and recording of their ideas. The researcher observations and participant feedback in our pilot study indicates that this approach immediately motivates users to explore varied design options in the given possibility space. The current proof-of-concept study indicates the value of the MindSculpt approach, and lays the groundwork for approaching more complex design scenarios and longer-term interactions between users and the system. We believe that MindSculpt opens exciting new avenues for the application of BCI to design research and practice.

## 5.1.   Limitations

Currently, MindSculpt is limited by the geometry generation algorithm. While linear combinations of four geometries provided many fascinating outputs (Figure 8), increasing the number of base shapes would allow the system to much more closely approximate the many possible geometries that an architectural designer would want to explore. More work is needed to determine how much variation in a user's brain patterns can be expected from day to day, and whether or not it may be possible for MindSculpt to accommodate such variations and adapt closer to a user over time. Finally, the current pilot study relied strongly on user feedback to evaluate the viability of the tool. While the strong positive reception and interest in using MindSculpt is a valuable indication of the system's potential, more work is needed to consider objective measures such as the visual similarity between user sketches, quantity and diversity of the MindSculpt output, as well as the practical effectiveness of integrating MindSculpt into actual design workflows.

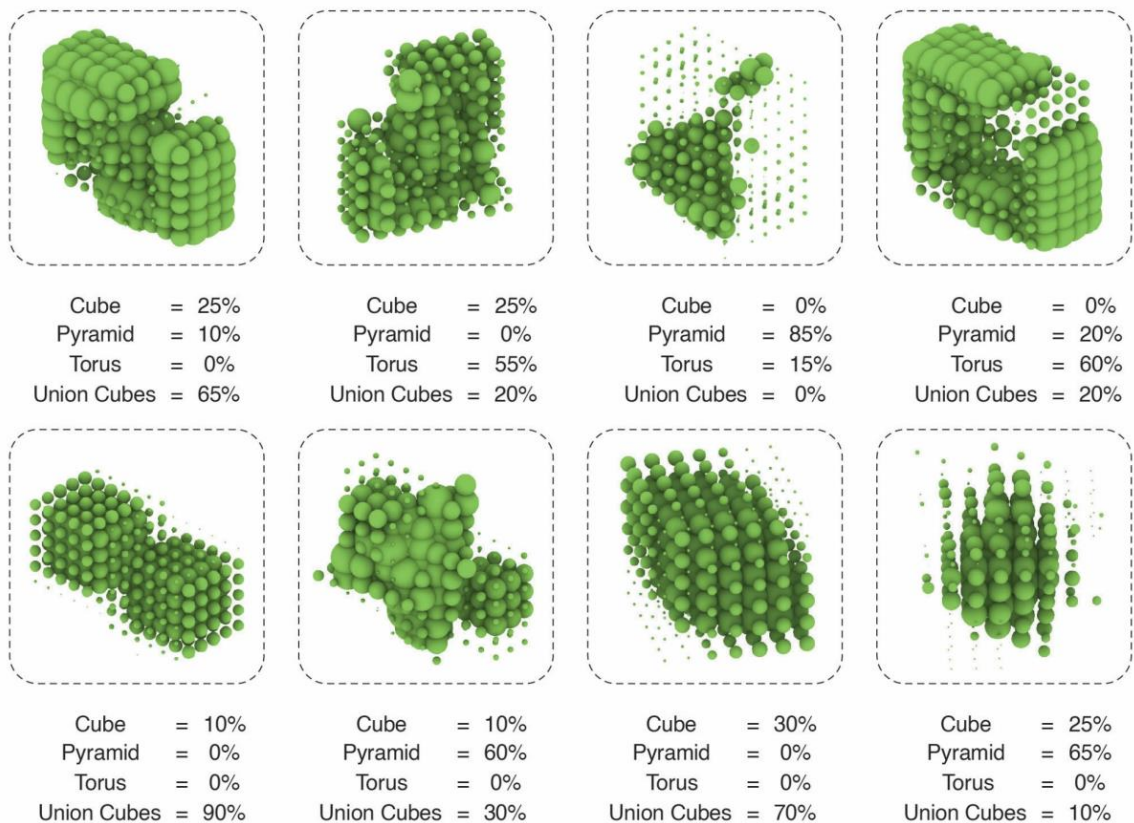

**Figure 8. A Wide Spectrum of Geometries Generated by MindSculpt.**

## 5.2.   Future Directions

To expand the possible design space of MindSculpt, we expect more fundamental studies that help us understand the association between the physiological indicators and design intentions including but not limited to EEG.

While MindSculpt provides a fascinating new BCI modality in design, it is not likely that the traditional mouse and keyboard paradigm, or traditional sketching and prototyping, will soon disappear. The most likely path forward in design will be some combination of BCI with these more traditional interfaces. Thus, future work in this area will need to evaluate the optimal uses of BCI during the design process. What design tasks are best implemented through BCI, and how can the use and output of these interfaces best be integrated with other aspects of the workflow? One approach to this question may emerge from the current strength of BCI in offering a high degree of control and fidelity for binary operations. Researchers may consider mapping BCI to commands that manipulate forms. For example, BCI could be used to choose between a "Boolean Union" vs. "Boolean Difference" relationship while other interface modalities are used to conduct "selection" or "changing position".

Making mistakes and encountering situations when design outcomes fail to match the original intention is expected during design processes. When this happens, however, designers need to be able to readily correct those mistakes or discover new directions embedded within them. The participants in the current study expressed some frustration when the outcome of the MindSculpt system diverged from their intent and could not be readily brought into line. Future work in this area will benefit from investigating the issue of error correction and non-BCI feedback in the system. A related issue is the steep learning curve of the BCI tools, which should be addressed by the development of clear guidebooks and interactive tutorials (Lotte and Camille 2015).

Future researchers in this area may also want to attend more closely to the paradigms of passive, reactive, and active BCI (Nijholt 2019, p.5-7). Reactive BCI requires users to attend to a stimulus (for example, a flickering segment of the displayed model) to select and alter that area. This type of BCI has higher accuracy but involves less voluntary thinking. Passive BCI monitors users' brain waves and triggers specific actions based on metrics such as stress levels. Though those paradigms are not precisely "doing by active thinking," they may have some potential for design applications. For example, one may imagine using passive BCI to evaluate and calibrate design options generated by active BCI, for example so that increasing user frustration would trigger a shift in the system's responses. As for active BCI, there are many potential opportunities to incorporate other types of neurological signals into the system's responses. Brain signals associated with various body movements of even emotional feelings could be used to create inspiring design processes. Future researchers should also focus more overtly on the value of BCI-based tools for designers who have motor disabilities or other limitations in using conventional WIMP-based computer interfaces. Attending specifically to this population will provide opportunities to evaluate and improve the overall utility of the technology and will help to expand its adoption.